\documentclass[reprint,
 superscriptaddress,
 amsmath,
 amssymb,
 nofootinbib,
 aps,
 pra,
]{revtex4-2}
\usepackage[utf8]{inputenc}
\usepackage{graphicx}
\usepackage{dcolumn}
\usepackage{bm}
\usepackage{natbib}
\usepackage{dirtytalk}
\usepackage{csquotes}
\usepackage{textgreek}
\usepackage[caption=false, position=bottom, justification = justified, singlelinecheck = false]{subfig}
\usepackage{adjustbox}
\usepackage{placeins}
\usepackage{csquotes}
\usepackage{amsfonts}
\usepackage{booktabs}

\begin{document}

\title{Self-strain suppression of the metal-to-insulator transition in \mbox{phase-change oxide devices}}

\author{Nicolò D'Anna}
\email{ndanna@ucsd.edu}
\affiliation{University of California San Diego, La Jolla, CA 92093, USA}
\author{Nareg Ghazikhanian}
\affiliation{University of California San Diego, La Jolla, CA 92093, USA}
\author{Erik S. Lamb}
\affiliation{University of California San Diego, La Jolla, CA 92093, USA}
\author{Edoardo Zatterin}
\affiliation{ESRF - The European Synchrotron, 71 Avenue des Martyrs, 38000 Grenoble, France}
\author{Mingze Wan}
\affiliation{University of California San Diego, La Jolla, CA 92093, USA}
\author{Ashley Thorshov}
\affiliation{University of California San Diego, La Jolla, CA 92093, USA}
\author{Ivan K. Schuller}
\affiliation{University of California San Diego, La Jolla, CA 92093, USA}
\author{Oleg Shpyrko}
\affiliation{University of California San Diego, La Jolla, CA 92093, USA}

\begin{abstract}
\textbf{Quantum materials exhibiting phase transitions which can be controlled through external stimuli, such as electric fields, are promising for future computing technologies beyond conventional semiconductor transistors.
Devices that take advantage of structural phase transitions have inherent built-in memory, reminiscent of synapses and neurons, and are thus natural candidates for neuromorphic computing.
Of particular interest are phase-change oxides, which allow for control over the metal-to-insulator transition.
Here, we report \mbox{X-ray} nano-diffraction structural imaging of micro-devices fabricated with the archetypal phase-change material vanadium sesquioxide (V$_2$O$_3$). The devices contain a Ga ion-irradiated region where the metal-to-insulator transition critical temperature is lowered, a useful feature for controlling  neuron-like spiking behavior.
Results show that strain, induced by crystal lattice mismatch between the pristine and irradiated material, leads to a suppression of the metal-to-insulator-transition. Suppression occurs within the irradiated region or along its edges, depending on the defect-distribution and the size of the region. 
The observed self-straining effect could extend to other phase-change oxides and dominate as device dimensions are reduced and become too small to dissipate strain within the irradiated region. 
The findings are important for phase engineering in phase-change devices and highlight the necessity to study phase transitions at the nanoscale.}
\end{abstract}

\maketitle

\begin{figure*} 
\centering   
    \includegraphics[width=\linewidth]{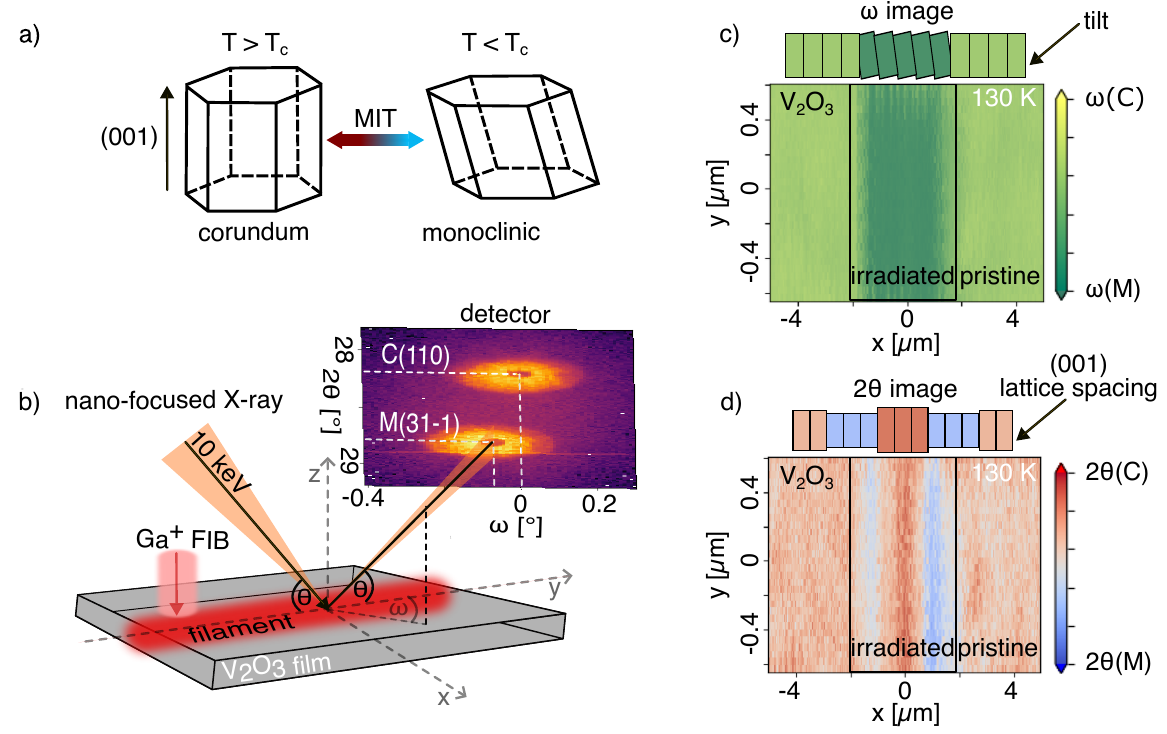} 
\caption{\textbf{\mbox{X-ray} nanoprobe imaging experiment.} 
    (a) Illustration of the crystal structure above (corundum) and below (monoclinic) the metal-to-insulator transition (MIT) temperature T$_c$.
    (b) Schematic of a V$_2$O$_3$ device with Ga irradiated region in red. 
    Specular reflection is recorded by the vertical axis of a 2D detector placed at an angle $\theta_D=2\theta$ with respect to incident \mbox{X-rays} with angle $\theta$ (with respect to the surface). The detector's horizontal axis measures the $\omega$ angle of the diffraction as defined in the figure. On the detector are shown the low temperature monoclinic M(31$\bar{1}$) and high temperature corundum C(110) Bragg peaks.
    (c) Typical image of the Bragg peak's $\omega$ center of mass throughout a device at 130~K, with the irradiated region visible as a low tilt region (dark green, illustrated above).
    (d) Corresponding image of the Bragg peak's $2\theta$ center of mass, with the irradiated region showing enhanced transition to the C(110) phase in the center and sidebands of suppressed out-of-plane lattice compression.
    }
\label{fig_setup}
\end{figure*} 

\noindent\textbf{1. Introduction}\\
Utilizing phase transitions for functional devices is a promising avenue for a variety of novel and energy efficient technologies, such as neuromorphic computing \cite{markovic2020physics,zhang2020understanding}, fast memory \cite{son2011excellent,park2017multi}, and optical switches \cite{liu2012terahertz}.
In the scope of neuron- and synapse-inspired neuromorphic computing, key requirements include integrated memory at the transistor level and neuron-like stochastic spiking behavior \cite{markovic2020physics}, both of which are found in functional metal oxides, including vanadium oxides \cite{mcwhan1970metal,zylbersztejn1975metal,mcwhan1973metal,adda2022direct,rischau2024resistive}, LSMO \cite{salev2024local,salev2021transverse,chen2024voltage,chen2024electrical}, LSCO \cite{zhang2020understanding,lengsdorf2004pressure}, NbO$_2$ \cite{pickett2013scalable,gao2017nbox}, and SmNiO$_3$ \cite{shi2013correlated,ha2013electrostatic,rischau2025synaptic}.
In these materials, the metal-to-insulator transition (MIT) can be triggered via electric and electrochemical gating in addition to temperature \cite{sawa2008resistive,li2018review}. 
In devices, sufficiently large electric fields can induce resistive switching: a locally induced transition into the metallic phase that typically occurs through the formation of a percolating conductive filament \cite{luibrand2023characteristic,duchene1971filamentary}. However, nanoscale parameters such as grain sizes, local stoichiometry, and intrinsic defects can cause large device-to-device variability \cite{shabalin2020nanoscale,del2018challenges,del2017electrically}. Nanoscale structural and electronic stochasticity can also lead to cycle-to-cycle variability \cite{cheng2021inherent}.

\begin{figure*} 
\centering   
    \includegraphics[width=\linewidth]{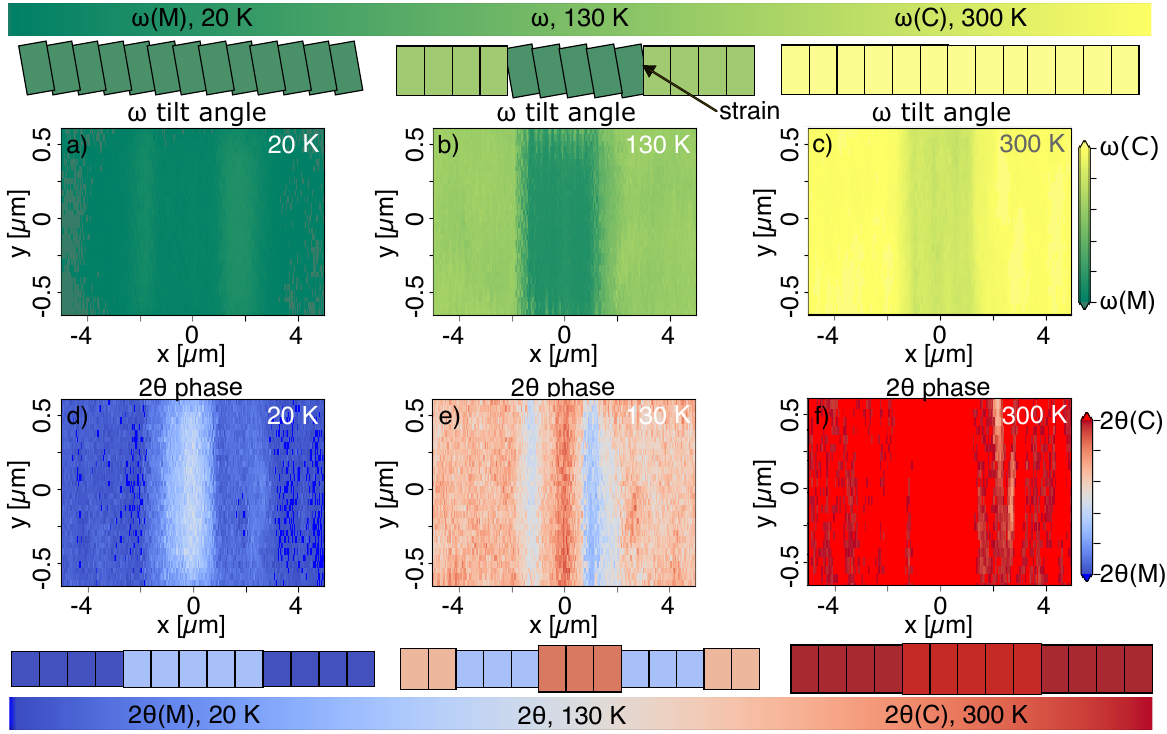} 
\caption{\textbf{Self-strain suppression of the metal-to-insulator transition.}
    Nanoprobe imaging of tilting of crystallographic planes ($\omega$) and out-of-plane lattice spacing ($2\theta$) for three temperatures across the MIT, in figure (a)-(c) and (d)-(f), respectively. The device has a 16~kV irradiated region.
    At 20~K $\ll$ T$_c$ ((a) and (d)), the device is in the low temperature monoclinic M(3,1,$\bar{1}$) phase with tilt $\omega$(M) and lattice spacing $2\theta$(M).
    At 130~K $\approx$ T$_c$ ((b) and (e)), the pristine V$_2$O$_3$ has reduced tilt and out-of-plane lattice spacing, half-way to the corundum C(110) phase 2$\theta$(C) and $\omega$(C). In the irradiated region tilting retains $\omega$(M) angle and induces strain in its edges which suppresses the MIT transition in the lattice spacing 2$\theta$.
    At 300~K $\gg$ T$_c$ ((c) and (f)), the pristine and irradiated regions are mostly in the C(110) phase 2$\theta$(C) and $\omega$(C), with the irradiated region retaining a $\sim$20\% tilt.
}
\label{fig_MIT_maps}
\end{figure*} 

Currently, there is an effort to improve control over conductive filament formation and increase performance stability upon cycling \cite{hu2023vanadium,wu2008influence}.
This may be achieved by locally altering electronic and structural properties to create a preferential path for filament formation.
Among other approaches, improvement  in stability was shown by local focused ion beam irradiation of oxide films \cite{xiang2021applications}, including VO$_2$ and V$_2$O$_3$ \cite{ghazikhanian2023resistive,hofsass2011tuning}.
The MIT critical temperature T$_c$ is lowered by irradiation-induced disorder, where irradiation energy controls induced defect concentration. Thus, transition to the metallic phase is facilitated in the irradiated region, allowing localized control of filament formation.
Device improvement, in terms of cycling stability and signal-to-noise ratio, depends on irradiation energy and dimensions \cite{xiang2021applications}.
In VO$_2$ and V$_2$O$_3$ the amount of Ga or He irradiation controls how much T$_c$ is lowered \cite{ghazikhanian2023resistive}.
To find the optimal fabrication parameters, a better understanding of the consequences of ion irradiation in oxide thin-films, at the micro- and nano-scale, is necessary.

Here, we use \mbox{X-ray} nano-diffraction to image the structural phase in Ga irradiated V$_2$O$_3$ devices at temperatures across the MIT. 
V$_2$O$_3$ has a structural phase transition coupled to the MIT.
Our experiment shows that at temperatures near the MIT, structural phase mismatch between the more conductive irradiated and the more insulating pristine V$_2$O$_3$ leads to self-induced strain on the irradiated region, which in turn suppresses the MIT. The exact location of the MIT suppression depends on the defect concentration and irradiation energy and can be reversed: for the low-irradiation energy region measured here, the MIT is suppressed within the center of the irradiated region, while for higher irradiation energies, it is suppressed along the edges. 
Thus, due to self-induced strain, the structural phase across irradiated regions is non-uniform and highly dependent on defect concentration (controlled by the irradiation energy). This highlights the importance of nano- to micro-scale structural effects in phase-change \mbox{functional devices.}
\mbox{Self-straining} should be expected when electrically switching most phase-change materials, possibly leading to physical limits on device dimensions.

\noindent\textbf{2. Results and Discussion}\\
In this work, three V$_2$O$_3$ micro-devices containing $\sim$1~$\mu$m wide Ga irradiated regions were imaged with \mbox{X-ray} nano-diffraction at various temperatures across the MIT upon warming from T$\ll$T$_c$, as illustrated in Fig.~\ref{fig_setup}b.
Each subsequent device was irradiated with a two-fold increase in ion energy over the same area, \textit{i.e.}, the irradiated regions in device \#1, \#2, and \#3 were exposed to a Ga beam with current 1~pC/$\mu$m$^2$ at 4, 8, and 16~kV respectively, denoted 4, 8, and 16~kV henceforth.
T$_c$ in the exposed region is lower than in the pristine region of the devices, such that there is a temperature range were the irradiated region is more metallic than the pristine region, as visible in Fig.~\ref{fig_setup}d.

\begin{figure} 
\centering   
    \includegraphics[width=\linewidth]{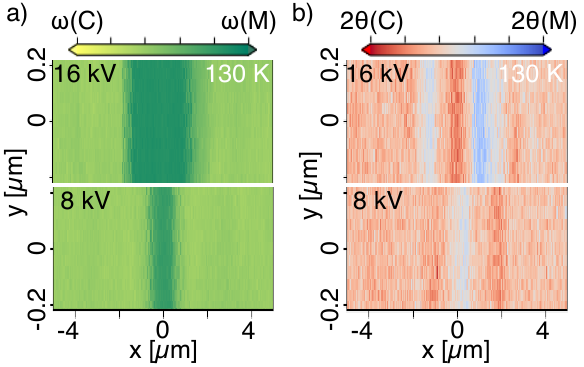} 
\caption{\textbf{Irradiation energy dependent self-straining.} Tilt (a) and out-of-plane lattice spacing (b) at 130~K $\approx$ T$_c$ for a device with 16, and 8~kV irradiated region, top and bottom, respectively. The same surface area was irradiated, however only a smaller part of the 8~kV irradiated region retains the tilt.
The lattice spacing shows suppression of the MIT ($2\theta$ remains closer to the low temperature C(110) phase $2\theta$(C)), on the edges of the 16~kV irradiated region and in the center of the 8~kV irradiated region.
}
\label{fig_sign_change}
\end{figure} 

In the \mbox{X-ray} nano-diffraction imaging, the diffraction at angles 2$\theta$ = 27.5$^{\circ}$ to 29.5$^{\circ}$, and $\omega$ = -0.6$^{\circ}$ to 0.6$^{\circ}$ is recorded on a two-dimensional detector. 
At photon energy 10~keV used here, the conductive phase corundum C(110) and the insulating phase monoclinic M(31$\bar{1}$) Bragg peaks of V$_2$O$_3$ are captured on the two-dimensional detector (see Fig.~\ref{fig_setup}b).
V$_2$O$_3$ is in the corundum (rhombohedral) paramagnetic metallic phase at room temperature and in the monoclinic antiferromagnetic insulating phase at low temperature (T$\lesssim$160~K), as illustrated in Fig.~\ref{fig_setup}a. 
The material exhibits a first-order phase transition, where the corundum structure transforms into monoclinic by expansion of the \textit{ab}-basal plane and contraction along the \textit{c} direction ([001], out-of-plane), such that the unit cell shears around the \textit{b}-axis \cite{dernier1970crystal}.
Thus, the C(110) peak is found at lower $2\theta$ and at higher $\omega$ angles than the M(31$\bar{1}$) peak.
For each position of the beam on the device, the Bragg peak's center of mass on the detector is computed, denoted $c_{2\theta}$ and $c_\omega$ for the center on the $2\theta$ and $\omega$ axis, respectively.
The conductive corundum C(110) peak position is taken to be the average position of $c_{2\theta}$ and $c_\omega$ in the pristine region of the device at 300~K, denoted $2\theta(C)$ and $\omega(C)$. Similarly, for the insulating monoclinic M(31$\bar{1}$) peak, the values $2\theta(M)$ and $\omega(M)$ are obtained at 20~K. The M(31$\bar{1}$) peak was found to have $\omega(M)$ consistently shifted to lower angles, indicating a preferential tilting direction in the devices.
Images of $c_\omega$ across the device, shown in Fig.~\ref{fig_setup}c, measure the tilt of the crystal lattice perpendicular to the incident beam direction (y-axis in Fig.~\ref{fig_setup}b). Such tilting is expected across the MIT in V$_2$O$_3$ \cite{kalcheim2020structural,shao2025x}.
Images of $c_{2\theta}$, shown in Fig.~\ref{fig_setup}d, measure the lattice parameter perpendicular to the surface ($Q_z$), as the $2\theta$-axis on the detector measures specular reflection.
Thus, $c_{2\theta}$ imaging shows the structural phase linked to the out-of-plane lattice spacing, while $c_\omega$ imaging shows tilt.

\begin{figure*} 
\centering   
    \includegraphics[width=\linewidth]{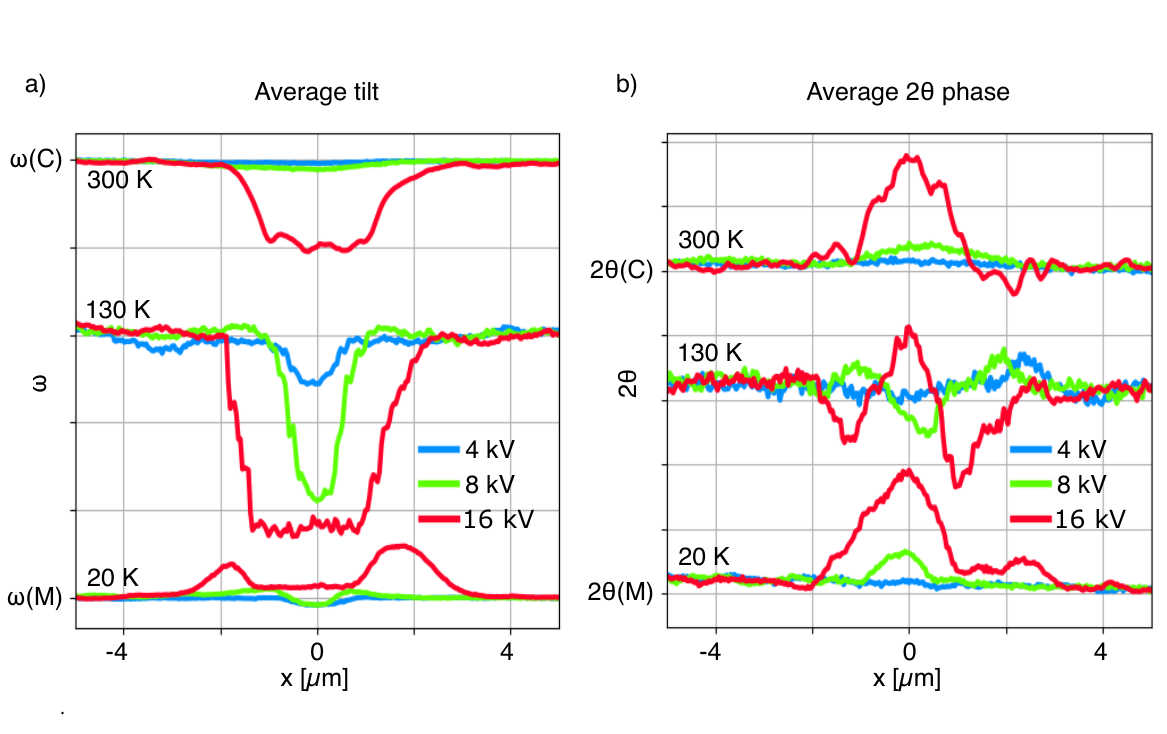} 
\caption{\textbf{Averages along y-axis.} 
    Plots obtained by averaging images as shown in Fig.~\ref{fig_MIT_maps} and Fig.~\ref{fig_sign_change} along the y-axis.
    \mbox{(a) Average} tilt across devices with regions irradiated at energies 4~kV (blue), 8~kV (green), and 16~kV (red), for three temperatures, 300~K $\gg$ T$_c$, 130~K $\approx$ T$_c$, and 20~K $\ll$ T$_c$.
    (b) Corresponding $2\theta$ average profiles.
    At 130~K, suppression of $2\theta$ change is strongest where the slope of $\omega$ change is largest. Indicating that MIT suppression is due to strain.
    }
\label{fig_line_cuts}
\end{figure*} 

Maps of the tilt $\omega$ and phase $2\theta$ in the device with the 16~kV Ga irradiated region are shown in Fig.~\ref{fig_MIT_maps} for three temperatures.
At the lowest temperature 20~K $\ll$ T$_c$, Fig.~\ref{fig_MIT_maps}a and .d, the phase and tilt are in the M(31$\bar{1}$) configuration, with the irradiated region lattice parameter more expanded (light blue region in Fig.~\ref{fig_MIT_maps}d).
Similarly, at room-temperature 300~K $\gg$ T$_c$, Fig.~\ref{fig_MIT_maps}c and .f, the device is in the C(110) phase, with a tilt mismatch in the irradiated region (light green region in Fig.~\ref{fig_MIT_maps}c).
For temperatures close to the pristine V$_2$O$_3$ MIT transition temperature T$_c$, T = 130~K $\approx$ T$_c$, there is the largest contrast between the Ga irradiated and pristine regions, Fig.~\ref{fig_MIT_maps}b and Fig.~\ref{fig_MIT_maps}e.
The irradiated region remains tilted in the low temperature M(31$\bar{1}$) position (dark green region in Fig.~\ref{fig_MIT_maps}b), which results in strain at the edges of the irradiated region. The strain is visible in Fig.~\ref{fig_MIT_maps}e as a suppression of the C(110) phase on the edges that remain in the more compressed and insulating M(31$\bar{1}$) phase (light blue). The center of the irradiated region remains tilted, but the out-of-plane direction is expanded indicative of the conductive C(110) phase as expected from the lower T$_c$ due to irradiation \cite{ghazikhanian2023resistive}.

The $\omega$ tilt at T = 130~K $\approx$ T$_c$ in the 8~kV and 16~kV Ga irradiated regions is shown in Fig.~\ref{fig_sign_change}a. Interestingly, for the device irradiated with 8~kV, the width of the region that retains the low temperature $\omega$ tilt is smaller than in the 16~kV device, despite the irradiated area being the same. Furthermore, the $\omega$ tilt value is identical for both 8 and 16~kV beams, suggesting an irradiation energy threshold effect.
The irradiation dose has a Gaussian profile, such that the center of the irradiated region has the highest dose. Thus, the region with irradiation above a certain threshold is larger in the 16~kV device than in the 8~kV device.
As a consequence of the narrower tilted region in the 8~kV device, strain suppresses the C(110) phase in the middle of the irradiated region. This is exactly the opposite to the effect on the 16~kV device, as visible on Fig.~\ref{fig_sign_change}b.

Two-dimensional maps presented in Fig.~\ref{fig_MIT_maps} and Fig.~\ref{fig_sign_change} are well summarized by averaging along the device \mbox{y-axis}, as is done in Fig.~\ref{fig_line_cuts}. The line averages are shown for three devices irradiated with 4~kV, 8~kV, and 16~kV (blue, green, and red data, respectively), at three temperatures, well above T$_c$ (T = 300~K $\gg$ T$_c$), near T$_c$ (T = 130~K $\approx$ T$_c$), and well below T$_c$ (T = 20~K $\ll$ T$_c$).
Figure~\ref{fig_line_cuts}.a shows the lattice tilt and Fig.~\ref{fig_line_cuts}b shows the phase.
For all irradiation energies, tilting is suppressed in the irradiated region (upon increasing temperature from 20~K) and higher irradiation energy leads to increased suppression. 
At T = 130~K $\approx$ T$_c$, the tilt mismatch between the irradiated and pristine regions is largest, leading to strain.
The effect of strain is visible in the $2\theta$ data, Fig.~\ref{fig_line_cuts}b, with the phase remaining more in the compressed insulating M(31$\bar{1}$) phase in the $\approx$ 1~$\mu$m wide irradiated region where the tilt has the most mismatch (largest slope in $\omega$ in Fig.~\ref{fig_line_cuts}a).
Interestingly, the $2\theta$ profiles for the three irradiation energies are similar at low and high temperatures, with the irradiated region being more in the metallic C(110) phase, while they differ substantially near T$_c$ due to strain. The 16~kV high irradiation energy device has two minima and one maximum in $2\theta$, while the 8~kV device has two maxima and one minimum, and the 4~kV device has no discernible $2\theta$ change in the irradiated region.

Nanoprobe imaging of pristine V$_2$O$_3$ observed the coexistence of multiple phases at the MIT, and a local strain induced suppression of the MIT \cite{shao2025x}.
Interestingly, in this work, the irradiated region retains tilt to higher temperatures than the pristine V$_2$O$_3$, at odds with a lower T$_c$ in the irradiated region. A possible explanation is that defects induced by the irradiation hinder the tilt relaxation, in agreement with the observation of transition temperature variation of up to 7~K induced by local strain and correlated with in-plane tilting \cite{shao2025x,zhu2016mesoscopic}. Here, we find that for irradiated V$_2$O$_3$ grown on (11-20) oriented sapphire, the dominant source of strain is at the interface between irradiated and pristine region. Global suppression of the MIT was observed in V$_2$O$_3$ as a consequence of strain induced by interaction with the substrate, which depends on substrate orientation \cite{kalcheim2020structural}, corroborating our results.
Similarly, strain was also observed to affects the MIT in VO$_2$ \cite{lee2017sharpened,yu2020controllable,hsieh2014evidence}.
It is also noteworthy that at \mbox{T $\approx$ T$_c$}, Fig.~\ref{fig_MIT_maps}b and .e, the center of the irradiated region has an expanded out-of-plane (001) lattice spacing and increased tilt, which are characteristics of the corundum and monoclinic phases, respectively. An explanation might be the formation of the paramagnetic insulating low-pressure phase of V$_2$O$_3$, which has the out-of-plane lattice parameter of the corundum phase and the in-plane lattice parameter of the monoclinic phase.

\begin{figure}[b] 
\centering   
    \includegraphics[width=\linewidth]{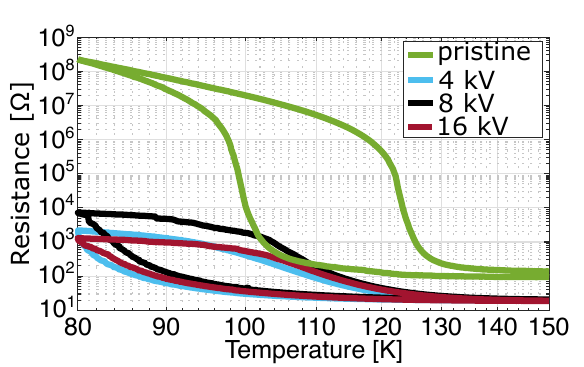} 
\caption{\textbf{Resistance vs temperature.} 
R vs. T measured prior to nanoprobe imaging for devices exposed to a Ga$^+$ beam current of 1 pC/$\mu$m$^2$ at various energies . At 8~kV irradiation energy, MIT suppression in the irradiated region leads to higher resistance.
}
\label{fig_RT}
\end{figure}

Results presented here are obtained by a single temperature cycle, starting at 20~K and warming to 300~K, such that the V$_2$O$_3$ starts in the insulating M(31$\bar{1}$) phase and ends in the conductive C(110) phase. This is relevant for applications where devices are kept in the insulating phase and locally switched to the conductive phase by an applied electric field.
Electrical switching creates a conductive filament, which will be sensitive to the phase tilt mismatch and induced strain observed in this work, warranting for similar measurements while electrically switching devices.
Resistance as a function of temperature for the devices studied here is shown in Fig.~\ref{fig_RT}, where the device with irradiation energy 8~kV has higher resistance than the other irradiated devices due to the observed suppression of the MIT (Fig.~\ref{fig_sign_change}).
Owing to V$_2$O$_3$'s strong conductivity hysteresis in temperature cycles \cite{del2017electrically}, it would also be of interest to expand these results to include a full temperature cycle starting from room temperature. In particular, to see whether the $\omega$ and $2\theta$ mismatch between irradiated and pristine V$_2$O$_3$ at 300~K are a result of the temperature cycle or is already induced by irradiation at 300~K.
Also of interest would be the addition of a fluorescence detector to image the distribution of Ga ions in the substrate and deduce the exact irradiation profiles \cite{masteghin2024benchmarking,DAnna_Xray_fluorescence}.
In addition, 3D Bragg coherent diffraction and ptychography imaging would be well suited to obtain additional information about defect locations and their effect on strain and phase. For example, defects might be responsible for the irradiated region's delayed relaxation to the untilted position.
Finally, combination with local conductivity imaging techniques, such as scanning microwave impedance microscopy \cite{barber2022microwave}, would allow direct comparison between local structure and conductivity to verify whether the structural phase transition and the metal-to-insulator transition are coupled at the nanoscale and in the presence of strain.

In terms of applications, the results are relevant for devices where switching is facilitated by an irradiated region. 
Notably, the results demonstrate irradiation energy is not the only parameter to consider, due to self-straining at the MIT temperature.
The width of the irradiated region is critical for achieving enhanced conductivity. To avoid strain reaching the center—essential for maintaining the metallic phase—the region must be wide enough, as demonstrated by the 16~kV device in Fig.~\ref{fig_sign_change}. If too narrow, as in the 8~kV device (Fig.~\ref{fig_sign_change} and Fig.~\ref{fig_RT}), self-straining suppresses conductivity, producing the opposite of the intended effect. Although all irradiated regions in this study were irradiated with the same width and method, they differed in energy, meaning their nominal dimensions were constant. Thus, optimizing the irradiated width for each irradiation energy is essential to ensure stability and improved performance in phase-change devices.


\noindent\textbf{3. Conclusion}\\
In conclusion, temperature dependent nanoscale structural imaging of V$_2$O$_3$ devices containing irradiated regions showed self-straining at the regions' border.
The strain is due to in-plane tilt mismatch between the metallic and insulating phases of V$_2$O$_3$ at temperatures near the metal-to-insulator transition (MIT) critical temperature T$_c$.
These sub-micron sized regions with strong strain are observed to cause the suppression of the MIT and their location is dependent on irradiation energy. In particular, if the irradiated region is too narrow, strain can suppress the MIT across the whole region and the opposite of the sought-after effect is achieved.
Such self-straining should be expected in all phase-change oxides, whereupon electrical switching forms a collection of metallic phase domains within an insulating matrix, creating boundaries of phase mismatch. Importantly, it could be a strong physical constraint to device miniaturization, posing a risk for technological applications.
Thus, for practical devices, self-straining effects in irradiated phase-change oxides have to be further studied and taken into account. 

\noindent\textbf{4. Experimental Section}\\

\textit{Sample Preparation:}
33-nm thick V$_2$O$_3$ films were grown on (1 1 -2 0)-oriented A-cut Al$_2$O$_3$ substrates via RF magnetron sputtering using a V$_2$O$_3$ target in a 7.9~mTorr Ar atmosphere and at 660~$^{\circ}$C substrate temperature. After the growth, the samples were thermally quenched at a rate of $\sim$90~$^{\circ}$C min$^{-1}$ to help preserve the correct oxygen stoichiometry and allow for dramatic improvement of electronic properties \cite{trastoy2018enhanced}. (100~nm~Au)/(20~nm~Ti) electrodes were fabricated in a 5~$\mu$m $\times$ 20~$\mu$m two-terminal geometry using standard photolithography techniques and e-beam evaporation. After the as-grown samples had been characterized, a 1~$\mu$m wide strip of Ga$^+$ ions bridging the electrodes across the entirety of the device was irradiated using a focused Ga$^+$ ion beam with a fluence of 6.2$\cdot$10$^{14}$ Ga ions/cm$^2$ and various accelerating voltages using a commercial scanning electron microscope.

\textit{X-ray nano-diffraction:}
The \mbox{X-ray} nano-diffraction experiments were conducted at the ID01 beamline at the European Synchrotron Radiation Facility (ESRF) \cite{leake2019nanodiffraction}. The setup used a Fresnel zone plate with 300~$\mu$m diameter and 60~nm outer zone width, focusing the beam to 65~nm at 10~keV. Diffraction was measured with a \mbox{MAXIPIX}, an in-house developed detector with 516$\times$516 pixels and pixel size 55~$\mu$m, placed at 4~m from the sample at an angle 2$\theta$=28.5$^\circ$.
The nano-diffraction data are available on the ESRF data portal \cite{ID01_data}.

\textit{Electrical Measurements:}
Electrical transport measurements were performed in a Lakeshore probe station using a Keithley 2450 source meter, Fig.~\ref{fig_RT}.

\begin{acknowledgments}
We acknowledge the European Synchrotron Radiation Facility (ESRF) for provision of synchrotron radiation facilities under proposal number HC-5367 and we would like to thank Edoardo Zatterin for assistance and support in using beamline ID01. 
This research was supported by the Quantum Materials for Energy Efficient Neuromorphic Computing, an Energy Frontier Research Center funded by the US Department of Energy (DOE), Office of Science, Basic Energy Sciences, under Award DE-SC0019273. This work was performed in part at the San Diego Nanotechnology Infrastructure (SDNI) of UCSD, a member of the National Nanotechnology Coordinated Infrastructure, which is supported by the National Science Foundation (Grant ECCS-2025752).
\end{acknowledgments}

\bibliography{biblio}		
\end{document}